\documentclass[conference]{IEEEtran}
\IEEEoverridecommandlockouts
\usepackage{my-conference-style}
\usepackage{custom-math}
\usepackage{cite}
\usepackage{amsmath,amssymb,amsfonts}
\usepackage{graphicx}
\usepackage{textcomp}
\usepackage{xcolor}
\usepackage{xurl} % Better URL formatting
\usepackage[inkscapelatex=false]{svg} % Supports SVGs for better image quality
\def\BibTeX{{\rm B\kern-.05em{\sc i\kern-.025em b}\kern-.08em
    T\kern-.1667em\lower.7ex\hbox{E}\kern-.125emX}}

\begin{document}

% Apply control to bib style
\bstctlcite{limit-authors-6} % Limits author names to 6 (Max. allowed per IEEE reference style guide)
\bstctlcite{disable-urls}
    
% Apply formatting to equation spacing
\setlength{\abovedisplayskip}{2pt}
\setlength{\abovedisplayshortskip}{2pt}
\setlength{\belowdisplayskip}{1pt}
\setlength{\belowdisplayshortskip}{1pt}

% Start document
\title{Efficient Probabilistic Assessment of Power System Resilience Using the Polynomial Chaos Expansion Method with Enhanced Stability\\ 
\thanks{The work is supported by Natural Sciences and Engineering Research Council (NSERC) Discovery Grant (NSERC RGPIN-2022-03236), Canada Research Chair Program, and Rubin \& So Faculty Scholar Award.}
}

\author{\IEEEauthorblockN{Aidan Gerkis\IEEEauthorrefmark{1} and Xiaozhe Wang\IEEEauthorrefmark{2}}
\IEEEauthorblockA{\textit{Department of Electrical and Computer Engineering} \\
\textit{McGill University}\\
Montréal, Québec, Canada \\
Email: \IEEEauthorrefmark{1}aidan.gerkis@mail.mcgill.ca, 
\IEEEauthorrefmark{2}xiaozhe.wang2@mcgill.ca}
}
\maketitle
\begin{abstract}
Increasing frequency and intensity of extreme weather events motivates the assessment of power system resilience. The random nature of power system failures during these events mandates probabilistic resilience assessment, but state-of-the-art methods are computationally inefficient. In this paper, an enhanced PCE method to quantify power system resilience based on the extended AC Cascading Failure Model (AC-CFM) model is proposed. To address repeatability issues arising from PCE computation with different sample sets, we propose a novel experiment design method. Numerical studies on the IEEE 39-bus system illustrate the improved repeatability and convergence of the method. The enhanced PCE method is then used to efficiently assess the system's resilience and propose adaptation measures.
\end{abstract}


\section{Introduction}\label{sc:intro}
Climate change has resulted in increasing frequency and intensity of extreme weather events, posing a severe risk to power systems. The ability of a power system to maintain its operation during these extreme events is encapsulated by the concept of resilience. Quantifying this resilience is an important step in the design of resilient power systems. 

To accomplish this task, Pantelli et al. \cite{Pantelli2017-ResM} proposed the $\flep$ metrics, quantifying resilience through the magnitude and rate of performance degradation and restoration during a resilience event, but only evaluated these metrics within a deterministic framework. However, since extreme events and the resulting power system failures are random in nature, resilience must be analyzed in a probabilistic one. To this end, Monte Carlo simulation (MCS) has been applied to quantify resilience (e.g. by \cite{Liu2017-MultipleTLMCS, Watson2020-ResWithRandomGen, Liu2024-ResAssessGSA}), but this method is computationally inefficient, requiring many simulations to accurately compute the probabilistic behaviour of resilience. To improve MCS' efficiency, scenario-based methods have been applied (e.g. by \cite{Wang2024-MIResAssess, Zhou2024-ResOHardening}). These methods determine ``typical" values for the random sources, approximating the system's average resilience with fewer model evaluations. However, they neglect the global resilience of the system. To overcome the computational challenge, Dobson et al. proposed a data-based method to assess cascading failures \cite{Dobson2012-PropOfCF}. While promising, this method is limited by the small amount of available data, restricting its ability to provide a global assessment of resilience.

To achieve computational efficiency while capturing the global distribution of uncertainty, recent works have applied the polynomial chaos expansion (PCE) method to power system problems such as transient stability \cite{Wang2021-DDSPCEforAPTC, Liu2022-PCCT-PCE} and under-voltage risk assessment \cite{Tan2024-GP-Risk}. While these methods have successfully modeled uncertainty in many power system problems, they have not yet been applied to assess power system resilience. Additionally, they have been shown to suffer from stability issues (i.e. the repeatability of the PCE approximation with different sample sets) \cite{Fajraoui2017-ED}, affecting their applicability to real-world problems.

To address these challenges, this paper applies the PCE method to efficiently quantify the probabilistic characteristics (i.e., mean, variance) of the resilience metrics based on an extended AC Cascading Failure Model (AC-CFM) \cite{Noebels-ACCFM}. In particular, an enhanced experiment design method is proposed to overcome the stability problem seen with the conventional PCE method, ensuring repeatable results across different sample sets.

\section{Resilience Event Modeling}\label{sc:res-model}
Power system failures during an extreme weather event typically begin with the failure of a few components. These initial failures can trigger a cascade of failures throughout the network, causing widespread outages \cite{Noebels-ACCFM}. Cascading failures occur when a change in the system's state causes a component to exceed its rated operating conditions, resulting in its disconnection from the network by a protection relay. 

\begin{figure}[ht]
   \begin{center}
        \captionsetup{width=1\columnwidth, belowskip=-0.5cm, font=small}
        \adjustbox{trim=0.9cm 0cm 1.25cm 0.4cm}{%
            \includesvg[width=1\columnwidth]{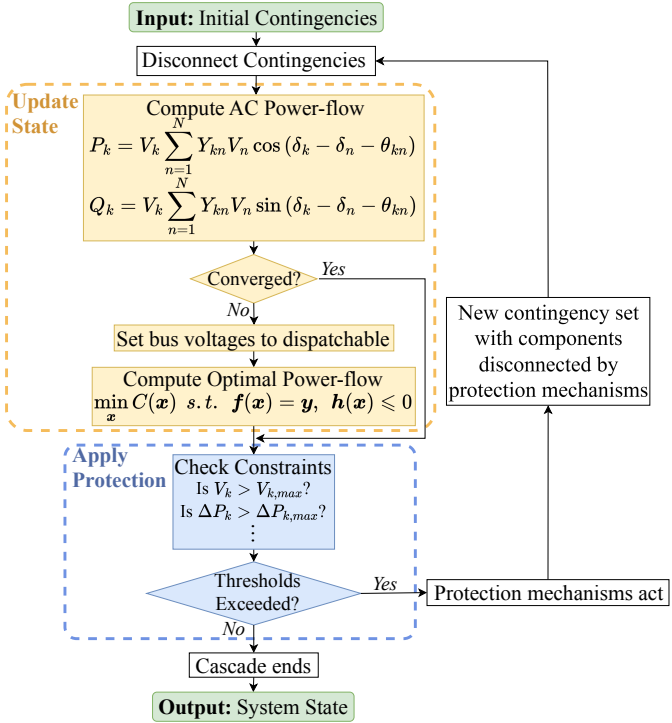}%
        }    
        \caption{The AC-CFM model proposed by Noebels et al. \cite{Noebels-ACCFM}.}
        \label{fig:ac-cfm}
    \end{center}
\end{figure}

AC-CFM, proposed in \cite{Noebels-ACCFM}, models cascading failures, calculating the system's response to a small set of initial contingencies. Firstly, the initial contingencies are disconnected from the network and the system state is updated by solving the AC power-flow. If this method fails to converge, the model relaxes the problem by converting all load buses to dispatchable and then applies the optimal power flow method to compute the resulting system state. Once convergence is achieved, the AC-CFM model checks if any components exceed their operating thresholds. Any that do are disconnected from the network to model the action of protection mechanisms, resulting in a new set of contingencies. This process repeats using the new set of contingencies until the system reaches a steady-state where no further protection mechanisms are triggered. Fig. \ref{fig:ac-cfm} summarizes the AC-CFM model, and the reader is referred to Noebels et al. \cite{Noebels-ACCFM} for a detailed discussion and validation.
%
%%%%% AC-CFM %%%%%%

The AC-CFM model computes a power system's response to a set of simultaneous contingencies, but extreme weather events typically consist of multiple non-simultaneous contingencies. To address this, we extend the AC-CFM model to a discretized model of a weather event where contingencies occur at hourly intervals, chosen to reflect typical protection actions, which are completed within one hour of an initial failure \cite{Dobson2012-PropOfCF}. The model input is a vector, $\boldsymbol{\tau}$, whose components $\tau_k$ represent the time at which the $k$-th component experiences a failure. At each time-step $t_i$, any components satisfying $\tau_k = t_i$ are set as the initial contingencies and the AC-CFM model calculates the resulting cascading failures and system state. This process is repeated at each time step to obtain the complete response to the extreme weather event. 

\begin{figure}[ht]
   \begin{center}
        \captionsetup{width=1\columnwidth, belowskip=-0.85cm, font=small}
        \adjustbox{trim=0cm 0.65cm 0cm 0.5cm}{%
            \includesvg[width=1\columnwidth]{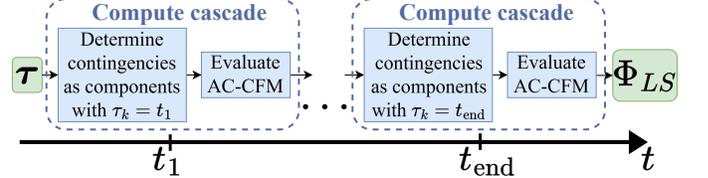}%
        }    
        \caption{The resilience model represented in \eqref{eq:fls-gen}.}
        \label{fig:res-model}
    \end{center}
\end{figure}

To evaluate the resilience of the system we apply the $\Phi$ metric \cite{Pantelli2017-ResM}, which quantifies the rate of change in a resilience indicator (i.e. a value quantifying the performance of the system) over time. We consider the load served indicator, whose value at time $k$ is defined as
\begin{align}
    P_{k}^{(\text{Served})} = \sum_{i\in\mathcal{L}}P_{(i, k)}^d\label{eq:ls-ind}
\end{align}
where $\mathcal{L}$ is the set of  load buses and $P_d^{(i, k)}$ is the power demanded at load bus $i$ at time $k$. To quantify resilience we apply the $\Phi$ metric, which measures the rate at which the load served decreases during a resilience event and is computed as
\begin{align}
    \flss = \frac{P_{(t_{\text{end}})}^{\text{Served}} - P_{(t_0)}^{\text{Served}}}{t_{\text{end}} - t_0}\label{eq:f}
\end{align}
where $t_0$ is the event start time and $t_{\text{end}}$ is the event end time \cite{Pantelli2017-ResM}. The resulting model may be expressed in the general form
\begin{align}
    \flss = \mathcal{M}(\boldsymbol{\tau})\label{eq:fls-gen}
\end{align}
where the input, $\boldsymbol{\tau}$, is a vector representing the failure time of each component in the power system. The discretized weather model, its inputs $\boldsymbol{\tau}$, and its output $\flss$, are summarized in Fig. \ref{fig:res-model}.
%
%%%%% Resilience Model %%%%%%
\vspace{-0.2cm}

The failure of a component during an extreme weather event is random, depending on the conditional probability of failure with respect to the weather state \cite{Pantelli2017-ResM}. We adopt the fragility curve to model random component failures. These curves specify failure probability as a function of weather state (e.g., wind speed) and are computed directly from historical outage data, thus modeling a wide range of failure mechanisms. Using the fragility curve, the failure time for a given component, $\tau_k$, can be modeled by some probability distribution function (PDF) $p_f^{(k)}$ and the elements of the vector $\boldsymbol{\tau}$ can be drawn by randomly sampling these PDFs so that $\tau_k \sim p_f^{(k)}$. In large transmission networks, the extreme weather events may affect only part of the network, so the vector $\boldsymbol{\tau}$ may only represent a subset of all components.

 While the mapping $\mathcal{M}$ is deterministic in nature, the randomness in the input $\boldsymbol{\tau}$ will propagate to the output, making the $\flss$ metric stochastic. Thus, to assess $\flss$, its statistical parameters (i.e., moments and/or distribution) must be computed. In this work we aim to compute the metric's moments. The mean of $\flss$ quantifies an average resilience event's impact, while the standard deviation describes how a real event may differ from the average. Computing both mean and standard deviation is necessary to quantify the range of possible events. The most common method to accomplish this is MCS, which evaluates the moments by generating many samples of $\boldsymbol{\tau}$ and evaluating \eqref{eq:fls-gen} on each sample. However, this approach is computationally expensive, requiring many evaluations of \eqref{eq:fls-gen} to accurately compute the moments.
 
\section{Polynomial Chaos Expansions}\label{sc:pce}
To overcome the MCS method's inefficiency, PCE models of the $\flss$ metric can be leveraged to efficiently evaluate its moments.

\subsection{Polynomial Chaos Expansion Theory \& Computation}\label{sc:gen-pce}
\vspace{-0.2cm}
\textbf{PCE Theory:} The PCE method approximates the $\flss$ metric using a polynomial function of the form
\begin{align}
   \hat{\Phi}_{LS} = \sum_{i=1}^Nc_i\boldsymbol{\Psi}_i(\boldsymbol{\tau})\label{eq:flss-pce}
\end{align}
where $\boldsymbol\Psi_i$ are multi-variate polynomials orthonormal with respect to the distribution of the random variable inputs, and $c_i$ are deterministic coefficients \cite{Blatman-PHD}. If the function \eqref{eq:fls-gen} is continuous then \eqref{eq:flss-pce} converges to $\flss$ in the $L^2$-norm sense \cite{Blatman-PHD}.

\textbf{The Experiment Design:} To compute the PCE representation in \eqref{eq:flss-pce}, a set of random input samples, $\boldsymbol{\mathcal{X}} = [\boldsymbol{\tau}^{(1)},..., \boldsymbol{\tau}^{(N_S)}]^T$, called the experiment, is required. The process of selecting the $N_S$ samples is called experiment design and is accomplished via MCS or Latin-hypercube sampling (LHS) of the input. Experiment design via MCS randomly samples each component of the input vector according to its PDF, requiring many samples to accurately represent the input space. LHS improves the representation by dividing the input space into $N_S$ equiprobable hypercubes and randomly placing one sample in each hypercube, providing more uniform coverage of the input space with fewer samples. In either case, \eqref{eq:fls-gen} is then evaluated on each input sample, resulting in a vector of model outputs $\boldsymbol{\Phi}_{LS} = [\flss^{(1)},..., \flss^{(N_S)}]^T$, which is used to compute the PCE. 

\textbf{Constructing the Polynomial Basis:} Before calculating the coefficients $c_i$, the multi-variate polynomial bases functions must be defined. These polynomials, $\boldsymbol{\Psi}_{\boldsymbol{\alpha}}(\boldsymbol{\tau})$, are computed as the tensor product of univariate polynomials $\psi_k^{(i)}(\tau_i)$ of degree $k$ which are orthonormal with respect to the distribution of the $i$-th component of the random input vector $\boldsymbol{\tau}$,
%\vspace{-0.3cm}
\begin{align}
    \boldsymbol{\Psi}_{\boldsymbol\alpha}(\boldsymbol{\tau}) = \prod_{i=1}^M\psi_{\alpha_i}^{(i)}(\tau_i)\label{eq:mv-poly}
\end{align}
where $M$ is the dimension of $\boldsymbol{\tau}$ and the multi-index notation is used to define the multi-variate polynomials. For some common distributions, known families of univariate orthonormal polynomials exist, but for arbitrary distributions the orthonormal polynomials can be computed through the Stieltjes procedure \cite{Blatman-PHD}. To improve sparsity in the PCE, the $q$-norm basis truncation scheme is adopted, limiting the set of all bases $\mathcal{A}$ to only polynomials with total degree $p$ or less:
%\vspace{-0.1cm}
\begin{align}
    \mathcal{A}^q = & \left\{\boldsymbol\alpha\in\mathcal{A}\::\:\left(\sum_{i=1}^M\alpha_i^q\right)^{1/q} \leq p\right\}\label{eq:q-norm-basis}
\end{align}

\textbf{Calculating the Coefficients:} For a given bases set $\mathcal{A}^q$, the coefficient vector, $\hat{\mathbf{c}}$, is computed using the experiment, $\boldsymbol{\mathcal{X}}$, and model evaluations, $\boldsymbol{\Phi}_{LS}$, as
\begin{align}
    \hat{\mathbf{c}} = \min_{\mathbf{c}}\mathds{E}\left[(\mathbf{c}^T\mathbf{A} - \boldsymbol{\Phi}_{LS})^2\right]\label{eq:ols-prob}
\end{align}
where $\mathbf{A}$ consists of evaluations of the basis polynomials on $\boldsymbol{\mathcal{X}}$. Equation \eqref{eq:ols-prob} is a least-squares regression problem, so a solution vector $\hat{\mathbf{c}}$ may be obtained analytically. In practice, the Hybrid-LARS method \cite{Blatman-PHD} is used to compute a sparse PCE by iteratively expanding the PCE basis, selecting the bases polynomials with the highest correlation to the residual and updating the coefficients until the desired accuracy is achieved. For more details see Chapter 5 in \cite{Blatman-PHD}.
\vspace{-0.2cm}
\subsection{Post-Processing of the PCE}\label{sc:pce-processing}
\vspace{-0.1cm}
Once the PCE model is computed, the mean and variance of the $\flss$ metric can be computed exactly from its coefficients.
\begin{align}
    \mu_{\flss} &= \mathds{E}\left[\hat{\Phi}_{LS}\right] = c_{\boldsymbol{0}}\label{eq:pce-mean}\\
    \text{Var}\left[\flss\right] &= \mathds{E}\left[\left(\hat{\Phi}_{LS} - \mu_{\flss}\right)^2\right] = \sum\limits_{\boldsymbol{\alpha}\in\mathcal{A}^q\backslash\boldsymbol{0}}c_{\boldsymbol{\alpha}}^2\label{eq:pce-var}
\end{align}
The PCE can also compute the distribution of $\flss$ by evaluating \eqref{eq:flss-pce} on many input samples.

\vspace{-0.25cm}
\subsection{The Challenges in Experiment Design and Estimation Consistency}\label{sc:ed-cons}
\vspace{-0.2cm}
The random sampling used to design the experiment, $\boldsymbol{\mathcal{X}}$, means that PCEs built with two experiment designs will have different moment approximations. Consider Table \ref{table:cons-ex} as an example. Two PCEs were built with different experiments of the same size. The PCE built with $\boldsymbol{\mathcal{X}}_1$ accurately approximates the moments, indicating convergence, while the other does not. The ability of a PCE to reliably converge when computed using different experiments is termed stability and is a critical barrier to the applicability of PCEs. If the PCE accuracy is inconsistent, the resulting PCE becomes unusable.
%
%%%%%% Consistency - Example %%%%%%
\begin{table}[!ht]
  \captionsetup{belowskip=-0.2cm, font=small}
  \caption{Moment approximations obtained from two PCEs of the resilience model described in Section \ref{sc:case-study} built with different experiments with $N_S = 30$.}
  \label{table:cons-ex}
  \centering
  \begin{tabular}{ccccc}
    \toprule
    \textbf{Method}                    & $\mu$    & Err($\mu$) & $\sigma$  & Err($\sigma$) \\\midrule
    MCS                                & -186.12  & ---        & 44.22     & ---        \\
    PCE - $\boldsymbol{\mathcal{X}}_1$ & -182.73  & 1.83\%     & 43.47     & 1.69\%     \\
    PCE - $\boldsymbol{\mathcal{X}}_2$ & -267.60  & 43.78\%    & 641.63    & 1350.96\%  \\\bottomrule
  \end{tabular}
\end{table}
\vspace{-0.4cm}

\section{A Novel Experiment Design Method}\label{sc:ed}
\begin{figure}[!ht]
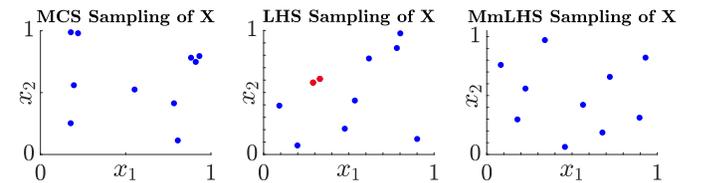

    \captionsetup{width=1\columnwidth, belowskip=-0.3cm, font=small}
    \centering
        \begin{subfigure}[t]{0.33\columnwidth}
        \captionsetup{belowskip=-0.2cm}
        \includesvg[width=1\columnwidth]{images/mcs-ex}%
        \caption{MCS sampling.} \label{fig:samplinga}
    \end{subfigure}%
    \hfill
    \centering
        \begin{subfigure}[t]{0.33\columnwidth}
        \captionsetup{belowskip=-0.2cm}
        \includesvg[width=1\columnwidth]{images/lhs-corner-case}%
        \caption{LHS sampling.} \label{fig:samplingb}
    \end{subfigure}%
    \hfill    
    \centering
        \begin{subfigure}[t]{0.33\columnwidth}
        \captionsetup{belowskip=-0.2cm}
        \includesvg[width=1\columnwidth]{images/maximin-lhs-ex}%
        \caption{MmLHS sampling.} \label{fig:samplingc}
    \end{subfigure}
    \caption{Sampling of a 2-D random vector with both components uniformly distributed in $[0, 1]$ using different methods.}
    \label{fig:sampling}
\end{figure}

%\vspace{-0.2cm}
To address the challenges in stability, a novel experiment design method, the Maximin-LHS method, is proposed. The conventional MCS method described in Section \ref{sc:gen-pce} may not sample the input space uniformly (see Fig. \ref{fig:samplinga}). As a result, PCEs computed using an MCS experiment design require many samples, $N_S$, to obtain an acceptable estimation accuracy. The LHS method improves this by attempting to uniformly cover the input space with fewer samples, though it does not guarantee uniformity (see red dots in Fig. \ref{fig:samplingb}).
%
%%%%%% Sampling Examples %%%%%%
\vspace{-0.4cm}

To obtain a more uniform covering of the input space, the Maximin sampling method \cite{Morris1995-MaximinLHS}, which designs an experiment by placing $N_S$ points so that the minimum distance between any two points is maximized, was introduced. Mathematically a Maximin experiment satisfies
\begin{align}
    \boldsymbol{\mathcal{X}}_{Mm} = \max_{\boldsymbol{\mathcal{X}}}\Bigl\{\min_{\mathbf{x}_1,\mathbf{x}_2\in\boldsymbol{\mathcal{X}}}||\mathbf{x}_1 - \mathbf{x}_2||_2\Bigr\}\label{eq:maximin}
\end{align}
One appeal of the Maximin experiment design is that it is D-optimal, implying that the information stored in the experiment is maximized \cite{Morris1995-MaximinLHS}.

\begin{figure}[ht]
   \begin{center}
        \captionsetup{width=1\columnwidth, belowskip=-0.8cm, font=small}
        \adjustbox{trim=1.2cm 0.2cm 1.2cm 0.4cm}{%
            \includesvg[width=0.9\columnwidth]{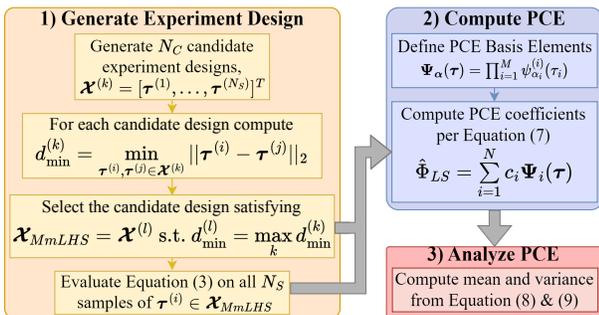}%
        }    
        \caption{The PCE uncertainty quantification method utilizing MmLHS experiment designs.}
        \label{fig:uq-fw-mmlhs}
    \end{center}
\end{figure}

While the Maximin method is theoretically sound, in practice it often results in experiments with sample points clumped on the input domain's boundaries. To overcome this, Morris et al. proposed the Maximin-LHS (MmLHS) method \cite{Morris1995-MaximinLHS} to design an experiment that satisfies \eqref{eq:maximin} and forms a Latin hypercube. Such a design will share the LHS designs' uniform covering properties while preventing clumping and being D-optimal \cite{Morris1995-MaximinLHS}. Although algorithms exist to compute MmLHS designs (e.g. the simulated annealing algorithm \cite{Morris1995-MaximinLHS}), they are computationally complex and assume a uniform distribution of the random vector. Thus, we propose a new algorithm for finding approximate MmLHS designs:
\begin{enumerate}
    \item Generate $N_{\text{C}}$ candidate LHS designs, $\mathcal{X}^{(i)}$, of size $N_S$.
    \item For each candidate design, compute %for each candidate design 
    the minimum Euclidean distance, $d_{\text{min}}^{(i)} = \min_{\mathbf{x}_1,\mathbf{x}_2\in\boldsymbol{\mathcal{X}}^{(i)}}||\mathbf{x}_1 - \mathbf{x}_2||_2$.
    \item Select the MmLHS experiment as $\mathcal{X}_{MmLHS} = \mathcal{X}^{(k)}\text{ s.t. } d_{\text{min}}^{(k)} = \max_i\left\{d_{\text{min}}^{(i)}\right\}$
\end{enumerate}
Fig. \ref{fig:samplingc} shows an MmLHS sampling obtained using this algorithm, clearly demonstrating an improved uniformity. This algorithm requires only the LHS design of $N_{\text{C}}$ experiments, so the time to compute an MmLHS design is only slightly longer than that to compute an LHS design. For example, when designing experiments with $N_S=25$ and six random inputs $t_{\text{LHS}} = 0.049s$ and $t_{\text{MmLHS}} = 0.15s$. Furthermore, since experiment design represents only a small fraction of the total PCE computation time (in the previous example $t_{\text{Total, LHS}}=911.41s$ and $t_{\text{Total, MmLHS}} = 905.82s$), the additional time required for the MmLHS design is negligible. The PCE computation method leveraging the MmLHS sampling method is summarized in Fig. \ref{fig:uq-fw-mmlhs}.
%
%%%%% UQ-Framework %%%%%%
\section{Case Study}\label{sc:case-study}
%\vspace{-0.2cm}
\subsection{Simulation Setup}
To evaluate the MmLHS experiment design method, PCEs were computed for the $\flss$ metric in the IEEE 39-bus system subject to a windstorm. The windstorm was assumed to occur over a 24 hour period and affect a small area within the power system covered by buses 12, 13, 14, 15, 16, 17, \& 20. It is assumed that component failures during the windstorm are caused by wind damage to transmission lines and towers, so contingencies are modeled at the branches connecting the affected buses. Failure times for each branch follow the probability distribution $p_f(t)$. To obtain this PDF, component failures during extreme windstorms in the Bonneville Power Administration \cite{BPA} were correlated with discretized wind speed data \cite{Merra2} and the resulting failure probabilities were calculated for each discretized wind speed value. The system's response to the failures was computed using the model described in Fig. \ref{fig:res-model}.

%\vspace{-0.2cm}
\subsection{Variance Calculation}
While PCE models can accurately predict the $\flss$ metric's behaviour in the distribution's body they struggle with behaviour in the tails. As a result, the standard deviation approximation obtained from the PCE represents the standard deviation in the distribution's body, not in the entire distribution \cite{Blatman-PHD}. In models with large outliers, this can cause significant discrepancies between the true standard deviation and the PCE's estimate. To better assess PCE accuracy, it is more appropriate to compare the approximated standard deviation (hereafter referred to as ``variance'') with an outlier-robust measure from the validation data. For this purpose the median absolute deviation, computed for a dataset $A$ as $\sigma = \text{median}\left(A - \text{median}(A)\right)$, was used to compute the true variance. 
%\vspace{-0.2cm}
\subsection{Stability Enhancement by the MmLHS Method}
\begin{figure}[!ht]
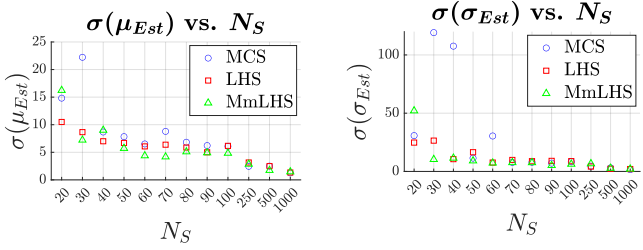

    \captionsetup{belowskip=-0.7cm, font=small}
    \centering
        \begin{subfigure}[t]{0.49\columnwidth}
        \captionsetup{belowskip=-0.6cm, font=small}
        \adjustbox{trim = 0cm 0.2cm 0cm 0.7cm}{%
            \includesvg[width=1\columnwidth]{images/var-mean-ns}%
        }
        \caption{Standard deviation of the mean estimate.} \label{fig:var-ns-a}
    \end{subfigure}%
    \hfill
    \centering
        \begin{subfigure}[t]{0.49\columnwidth}
        \captionsetup{belowskip=-0.2cm, font=small}
        \adjustbox{trim = 0cm 0.1cm 0cm 0.7cm}{%
            \includesvg[width=1\columnwidth]{images/var-var-ns}%
        }        
        \caption{Standard deviation of the variance estimate.} \label{fig:var-ns-b}
    \end{subfigure}
    \caption{The standard deviation of the moment approximations vs. $N_S$.}
    \label{fig:var-ns}
\end{figure}

To evaluate stability, experiments of different sizes $N_S$ were generated using the MCS, LHS, and MmLHS methods, with 25 unique experiments being generated for each $N_S$ and each method. Each point in Fig. \ref{fig:mean-ns-a} represents the average of the 25 mean approximations obtained from the PCE models computed with a specific sample size $N_s$ and sampling method. The same procedure was repeated to obtain the remaining plots in Fig. \ref{fig:mean-ns} and \ref{fig:var-ns}. The moment approximations were validated by comparison to the mean and variance computed via MCS, represented by the dashed lines in Fig. \ref{fig:mean-ns-a} \& \ref{fig:mean-ns-b}. A total of 10,000 MCS were required to obtain convergence in the mean and variance.
%
%%%%% Mean - Mean & Variance Approximation %%%%%%
\begin{figure}[!ht]
    \captionsetup{belowskip=-0.2cm, font=small}
    \centering
        \begin{subfigure}[t]{0.49\columnwidth}
        \captionsetup{belowskip=-0.2cm}
        \adjustbox{trim = 0cm 0.2cm 0cm 0.45cm}{%
            \includesvg[width=1\columnwidth]{images/mean-mean-ns}%
        }
        \caption{Mean of the mean estimate.} \label{fig:mean-ns-a}
    \end{subfigure}%
    \hfill
    \centering
        \begin{subfigure}[t]{0.49\columnwidth}
        \captionsetup{belowskip=-0.2cm}
        \adjustbox{trim = 0cm 0.2cm 0cm 0.45cm}{%
            \includesvg[width=1\columnwidth]{images/mean-var-ns}%
        }
        \caption{Mean of the variance estimate.} \label{fig:mean-ns-b}
    \end{subfigure}
    \caption{The mean of the moment approximations vs. $N_S$. The true moments are plotted in black.}
    \label{fig:mean-ns}
\end{figure}
%
%%%%% Standard Deviation - Mean & Variance Approximation %%%%%%

Fig. \ref{fig:mean-ns-a} shows that the mean approximations are consistently accurate across all methods, with no significant improvements from the MmLHS method. However, it can be observed from Fig. \ref{fig:mean-ns-b} that the variance approximation converges with smaller $N_S$ when using MmLHS, while MCS and LHS show decreases in accuracy at $N_S = 60$ and $N_S = \{70, 90\}$, respectively. Fig. \ref{fig:var-ns-a} demonstrates that the standard deviation of the mean approximations is consistently smaller with the MmLHS method for smaller $N_S$ ($N_S \in [50, 100]$). It is also observed from  Fig. \ref{fig:var-ns-b} that the standard deviation of the variance approximations decreases faster when using the MmLHS method, dropping to 10.32 when $N_S = 30$. After $N_S = 30$, the standard deviation for the MmLHS method remains low, indicating a stable approximation. In contrast, MCS and LHS show significant increases in standard deviation of the variance at $N_S = 60$ and at $N_S = 50$, respectively. 

These observations indicate that the MmLHS method obtains better stability in the PCE model and provides faster convergence for variance approximations, meaning that smaller experiment sizes can achieve accurate PCE models. For large $N_S$, all methods perform similarly, with the standard deviation of moment approximations, Fig. \ref{fig:var-ns-a} and \ref{fig:var-ns-b}, trending to 0. 
%\vspace{-0.cm}
\subsection{Insights from the Resilience Assessment Results}
To assess the power system's resilience to the windstorm, we examine the mean and variance approximations from the PCE model with the lowest computation error (Table \ref{table:fls-mom}). The mean provides an estimate of the expected rate of load-shedding in a typical failure scenario, which provides important guidance in the planning of adaptation measures (e.g., deployment of backup generation). However, the large variance indicates that actual load-shedding rates could significantly exceed the mean value. This observation highlights the importance of assessing the global behaviour of the $\flss$ metric, as designing response measures based only on the mean value may result in inadequate preparedness. To address this, the variance may be used to generate confidence intervals for the $\flss$ metric, enabling more robust decision-making. For example, the system operator may deploy backup generators or portable energy storage to provide additional generation at a rate of $-303.89$ MW/Hr (a lower bound on $\flss$ computed by the 3$\sigma$ rule) to handle higher-than-expected load shedding and enhance the system's resilience during the storm. The proposed method is scalable to larger power systems, although a larger experiment design ($N_S \approx 10\times M$) will be needed.

%
%%%%%% FLS - Moment Comparison %%%%%%
\vspace{-0cm}
\begin{table}[ht]
    \captionsetup{belowskip=-0.3cm, font=small}
    \caption{Estimated moments of the $\flss$ metric, obtained from a PCE model computed with the MmLHS method and $N_S = 70$.}
    \label{table:fls-mom}
    \centering
    \begin{tabular}{cccccc}
        \toprule
        \textbf{Model}    & $\mu\: [MW/Hr]$    & Err($\mu$) & $\sigma$ & Err($\sigma$) & $t_{\text{Total}}$ \\\midrule
        MCS               & -186.12  & ---        & 44.22    & ---           & 208933.87s   \\
        PCE               & -178.13  & 4.29\%     & 41.92    & 5.20\%        & 431.88s      \\\bottomrule
    \end{tabular}
\end{table}
\FloatBarrier
\vspace{-0.9cm}
\section{Conclusion}\label{sc:conc}
\vspace{-0.1cm}
In this paper, we integrated the MmLHS experiment design method with the PCE method. Numerical studies showed that MmLHS improved the PCE method's stability and sped-up convergence. The PCE method was then used to assess system resilience and insights from the model were discussed. Future work includes investigating scalability investigating other methods for improving consistency (i.e., importance-sampling), and applying the PCE method to design adaptation measures.
\vspace{-0.3cm}
\bibliographystyle{IEEEtran}
\bibliography{IEEEabrv, master-references}

\end{document}